\documentclass{elsart}

\usepackage{graphics}

\begin{document}

\begin{frontmatter}

\title{Solving the Master Equation for Extremely Long Time Scale Calculations}

\author{Hwee Kuan Lee, Yutaka Okabe}

\address{Department of Physics, Tokyo Metropolitan University, Hachioji, Tokyo 192-0397, Japan}

\author{X. Cheng, M. B. A. Jalil}

\address{Department of Electrical and Computer Engineering, National University of Singapore, Singapore 119260, Republic of Singapore}

\begin{abstract}
The dynamics of magnetic reversal process 
plays an important role in the design of
the magnetic recording devices in the long time scale limit.
In addition to long time scale, microscopic effects such as 
the entropic effect become important in magnetic nano-scale systems.
Many advanced simulation methods have been developed, but few have 
the ability to simulate the long time scale limit and to
accurately model the microscopic effects of nano-scale systems
at the same time.
We develop a new Monte Carlo method for calculating the dynamics 
of magnetic reversal at arbitrary long time.
For example, actual calculations were
performed up to $10^{50}$ Monte Carlo steps.
This method is based on microscopic interactions of many constituents
and the master equation for magnetic probability
distribution function is solved symbolically. 
\end{abstract}

\end{frontmatter}

\section{Introduction}
Studies on complex systems with meta-stable states are valuable in many
disciplines. For instance, aging phenomena affect the properties of materials,
magnetic relaxation processes determine the functions of magnetic recording media,
and the study of first order phase transitions and spinodal decomposition have
wide applications. Often, slow dynamics plays a central role in the 
understanding of these complex systems. 
When no analytic solutions are 
available, numerical simulations are the only tools to study these systems.
In a Monte Carlo simulation, the meta-stable states could 
manifest as apparently stable or ``equilibrium" states. Only after a very
long time the simulation could overcome the free energy barrier and go
into equilibrium. When using the standard Metropolis algorithm, this time scale
may become unattainable even 
with the fastest computer available. Hence advanced Monte
Carlo techniques such as the Absorbing Markov Chains 
algorithm~\cite{novotny}, which is a 
generalization of the N-fold way method~\cite{bortz}, and the Projection 
Method~\cite{kolesik} have been developed to address this issue. These methods 
have been used to study magnetic reversal~\cite{novotny1,munoz,brown}. In some
cases, where the difficulties of meta-stability are not too serious, 
conventional Monte Carlo methods are also 
used~\cite{hinzke,nowak1,nowak2,dittrich}. 

In this paper, we will present a new
method of calculating the dynamics of systems with meta-stable states 
by solving the master equation for magnetic probability distribution function
symbolically. We
illustrate our method in the context of magnetic reversal on the two
dimensional Ising square lattice.

\section{Method}
The Ising model is 
a basic model to study the phase transition, and has also been used as
a standard benchmark model for testing new 
algorithms\\
~\cite{novotny1,wangjs,ferrenberg1,wolff,ferrenberg2,lee1,oliveira,yamaguchi1,yamaguchi2,wang,wang1}. 
%
%
The Hamiltonian for the Ising model is
\begin{equation}
{\mathcal H} = -J \sum_{<i,j>} s_i  s_j - h \sum_{i} s_i = E -h M
\label{eq:hami}
\end{equation}
where $s_i = \pm 1$. The first summation is over nearest neighbors; the second 
summation is over all lattice sites. $J$ and $h$ are the exchange constant and
external field respectively. We define two symbols $E$ and $M$ as the energy 
of the exchange part and the magnetization part,
\begin{eqnarray}
\label{em}
E & = & -J \sum_{<i,j>} s_i  s_j \nonumber \\
M & = & \sum_{i} s_i
\end{eqnarray}
Ideally, to study the dynamics of magnetic reversal on the Ising model,
the microscopic master equation,
\begin{equation}
\frac{d P(\sigma,t)}{d t} = \sum_{\sigma '} 
                    \omega (\sigma | \sigma ')
                    P(\sigma ',t) -
                    \omega (\sigma ' | \sigma )
                    P(\sigma ,t) 
\label{eq:masterm}
\end{equation}
is to be solved. $\sigma$ denotes the microscopic
state of the system, $\omega (\sigma | \sigma')$
is the transition rate from $\sigma'$ to $\sigma$, and $P(\sigma,t)$ is the 
probability distribution of the state $\sigma$ at time $t$. For a lattice with
$N$ sites, Eq. (\ref{eq:masterm}) is a set of $2^N$ coupled differential equations
and solving them is infeasible. Instead the dynamics can be approximated by writing
the master equation in terms of the total magnetization $M$,
\begin{figure}
\begin{picture}(0,210)
\put(40,0){\scalebox{0.4}{\includegraphics{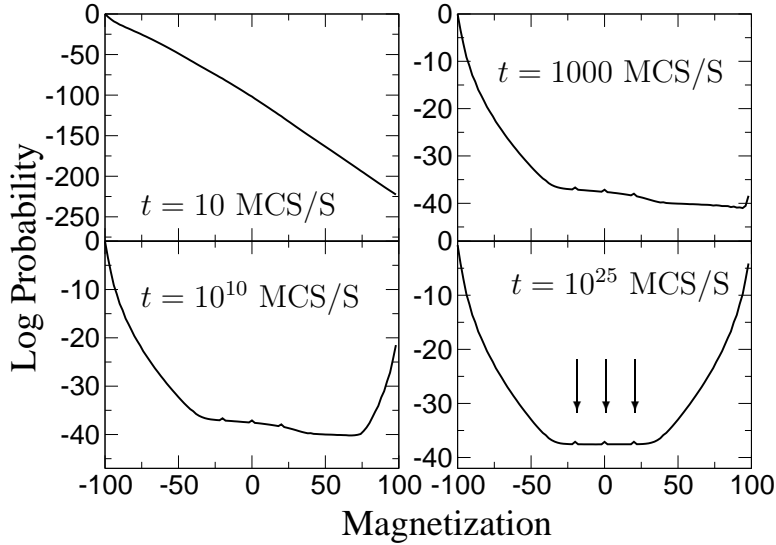}}}
\put(90,125){$t = 10$ MCS/S}
\put(225,175){$t = 1000$ MCS/S}
\put(90,90){$t = 10^{10}$ MCS/S}
\put(230,95){$t = 10^{25}$ MCS/S}
\put(255,70){\vector(0,-1){20}}
\put(266,70){\vector(0,-1){20}}
\put(277,70){\vector(0,-1){20}}
\end{picture}
\caption{Log of probability distributions plotted at $t=10,1000,10^{10}$ and
$10^{25}$ for $L=10, T = 0.44 T_c$ and $h=0$. Initial condition is
 $P(t=0,M) = \delta_{M,-100}$. At time $t=10^{25}$, the probability
distribution is almost indistinguishable from the equilibrium distribution. The
little spikes marked with arrows are due to band structure of the Ising 
model (Fig. \ref{fig:band}). Equilibrium probability distribution was 
generated using the Wang-Landau method up to correction factor of
log($f$) = $1.25 \times 10^{-8}$. Error bars are smaller than the thickness 
of the lines.}
\label{fig:prob}
\end{figure}
\begin{equation}
\frac{d P(M,t)}{d t} = \sum_{M '} 
                    \omega (M | M ')
                    P(M ',t) -
                    \omega (M ' | M )
                    P(M ,t) 
\label{eq:masterM}
\end{equation}
which is a set of $N+1$ differential equations.
The approximate form of the macroscopic transition rates $\omega(M|M')$ was
discussed by Lee $et. al.$~\cite{lee}.
We made identical approximations in the transition rates 
as the ``mean field dynamics 2" (MFD2) obtained by Lee $et. al.$~\cite{lee}. 
Other ways of obtaining the transition rates, such as
the Transition Matrix Monte Carlo method~\cite{swendsen}, may also be used.
Our main contribution in this paper is in the way of solving 
the macroscopic master equation (Eq. (\ref{eq:masterM})). 
Our results are accurate to within the validity of basic 
approximations in the transition rates. However, the gain in efficiencies and 
accuracies of our results are many orders of magnitude larger than most Monte 
Carlo methods.

As in Ref.~\cite{lee} the equilibrium distribution is used to calculate the 
transition rates. With this approximation, not all microscopic information is
available. The region of validity for this approximation will be discussed 
later.

An efficient Monte Carlo algorithm to calculate the density of states has been
recently proposed by Wang and Landau~\cite{wang,wang1}. We have used
the Wang-Landau~\cite{wang,wang1} algorithm to estimate the density of
states as a function of exchange energy and magnetization $g(E,M)$, with $E$ 
and $M$ defined by Eq. (\ref{em}).
For a given $\beta=1/k_{\mbox{\tiny B}} T$ and $h$,
the joint probability distribution 
$P^{eq}_{\beta,h}(E,M)$ is then obtained using Boltzmann weights as
\begin{equation}
P^{eq}_{\beta,h}(E,M) = g(E,M) \mbox{ exp}(- \beta [E - h M ]) / {\mathcal Z}(\beta,h)
\end{equation}
$\mathcal Z$ is the partition function given by,
\begin{equation}
{\mathcal Z}(\beta,h) = \sum_{E,M} g(E,M) \mbox{ exp}(-\beta [ E -  h M ])
\end{equation}
Finally, the equilibrium probability distribution of $M$ can be obtained,
\begin{equation}
P^{eq}_{\beta,h}(M) = \sum_{E} P^{eq}_{\beta,h}(E,M)
\end{equation}

For a lattice with $L^2=N$ sites, there are $N+1$ possible values of
magnetization. Then the master equation (Eq. (\ref{eq:masterM}))
can be written in a matrix form,
\begin{equation}
\frac{d \vec P(t) }{dt} = {\mathcal A} \cdot \vec P(t)
\label{eq:masterMat}
\end{equation}
where the $(N+1)$ dimensional vector $\vec P (t)$ is
\begin{equation}
\vec P (t) = \left(
   \begin{array}{c}
       P(M=-N,t) \\ \vdots \\ P(M=0,t) \\ 
       \vdots \\ P(M=+N,t) 
   \end{array} \right) 
\end{equation}
In this paper, we present a way to calculate the explicit solution to
Eq. (\ref{eq:masterMat}). We restrict
the transitions to $| \Delta M |$ = 2. This is a vital component of our work 
and we shall elaborate its implications and the conditions in which it is 
valid. Transitions which involve $|\Delta M|$ = 2 are all single spin flip 
moves or any move which flip $N$ spins from $+1$ to $-1$ and $N+1$ spins from
$-1$ to $+1$ (or vice versa). At low temperature and low field, we do not 
expect to observe many cluster flips or multiple spin flip transitions. Hence 
our assumption is valid at low temperature and low field. Incidentally, the 
process of magnetic reversal at low temperature and field is via the 
single-droplet process~\cite{langer1,langer2,rikvold,tomita}. This is also the domain
where the approximation to our transition rates is 
valid. At high temperatures, other magnetic reversal processes such as the 
multi-droplet process dominate and our approximation becomes not accurate in this
regime. Short time scale behavior becomes important at high temperatures
and other Monte Carlo methods may be used.
\begin{figure}
\begin{picture}(0,330)
\put(25,0){\scalebox{0.33}{\includegraphics{./10x10-domain.eps}}}
\put(80,205){\scalebox{0.33}{\includegraphics{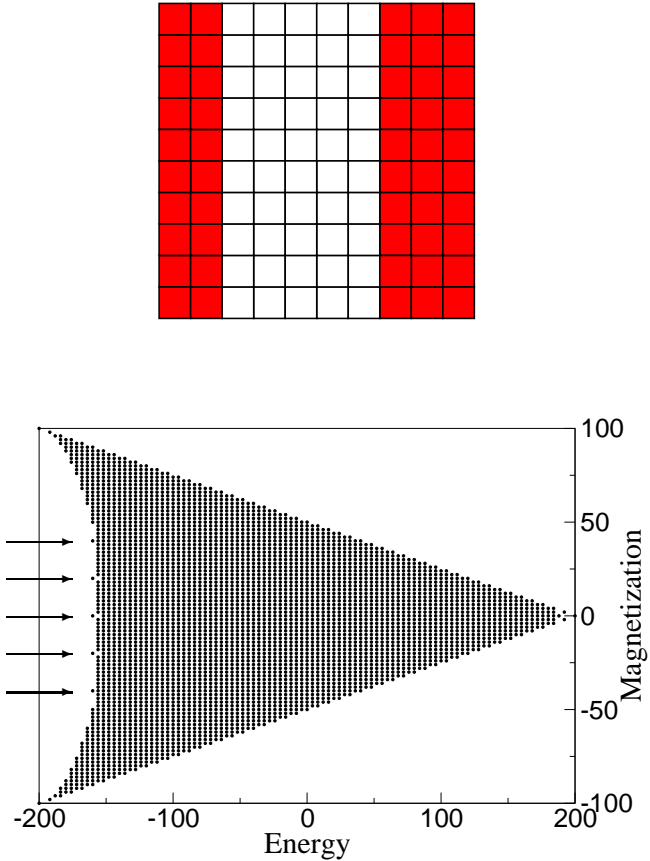}}}
\put(23,64){\vector(1,0){25}}
\put(23,78.5){\vector(1,0){25}}
\put(23,92.5){\vector(1,0){25}}
\put(23,107){\vector(1,0){25}}
\put(23,121){\vector(1,0){25}}
\end{picture}
\caption{The upper figure shows a band structure that 
is the origin of spikes observed 
in the probability distribution of Fig. \ref{fig:prob}. 
These are local
energy minimum configurations, and any single spin flip increases
the energy. The lower figure shows the domain of accessible state for $L = 10$ 
Ising model. The minimum energy state with $M = 0$ is $E = -160$ or 
$E = E_g + 4 L$, where $E_g$ is the ground state energy. Arrows indicate band 
structure states shown in the upper figure, and 
their exact density of states can also be trivially calculated.}
\label{fig:band}
\end{figure}

With the constraint $|\Delta M|$ = 2, the transition rate $\omega$ is 
written as
\begin{equation}
\omega ( M_j | M_i ) \approx \left\{
\begin{array}{cc} 
 P^{eq}(M_j) / [ P^{eq}(M_j)+ P^{eq}(M_i) ]  & | i-j | \leq 1 \\
0 & \mbox{otherwise} 
\end{array} \right. 
\label{eq:trans}
\end{equation}
if the Glauber rate is employed.
The Metropolis rate can also be used, and 
we compare the results obtained by both rates in Table \ref{tbl:err}.

For the transition rates given by Eq. (\ref{eq:trans}), the master equation 
(Eq. (\ref{eq:masterMat})) can be reduced into one linear differential equation
with constant coefficients and the analytic solutions for this equation is 
known~\cite{ince}. A procedure to derive the solution is given in an appendix. 
The general solution is of the form,
\begin{equation}
\vec P(t) = \sum_{i=0}^{N} \alpha_i \vec v_i \mbox{exp}(\lambda_i t)
\label{eq:pt}
\end{equation}
where $\lambda_i$ and $\vec v_i$ are the eigenvalues and eigenvectors of the
matrix ${\mathcal A}$ and $\alpha_i$ are constant coefficients determined by 
solving Eq. (\ref{eq:pt}) with initial conditions,
\begin{equation}
\vec P(0) = \sum_{i=0}^{N} \alpha_i \vec v_i 
\end{equation}
We used MATHEMATICA to calculate $\alpha_i$, $\vec{v}_i$ and $\lambda_i$ 
numerically, and $\vec{P}(t)$ is represented symbolically as a function of 
time. We have also made use of a MATHEMATICA function for setting arbitrary 
precisions and compared our results for calculations with different precisions.
With $\vec{P}(t)$ evaluated symbolically, other magnetic quantities, such as
$M(t)$, can be evaluated symbolically as well.

The initial conditions were set with all 
the spins in the $-1$ position and the external field 
always favors spin configurations with $s_i = +1$. 
Our simulations were performed on a dual 2 GHz PowerMac G5 PC, with over 95\% 
of the computing resources devoted to calculating the equilibrium density of 
states using the Wang-Landau method~\cite{wang,wang1}.

\section{Results}

All eigenvalues are real and negative except for the largest eigenvalue which 
is zero. The probability distribution can be re-written as,
\begin{figure}
\begin{picture}(0,200)
\put(30,0){\scalebox{0.35}{\includegraphics{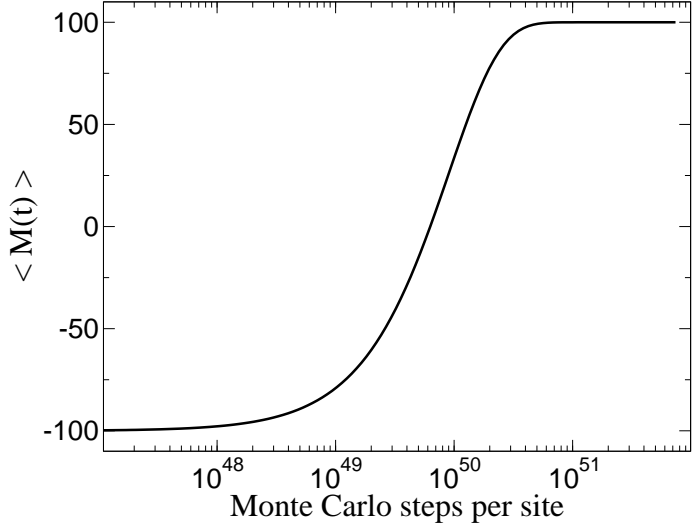}}}
\end{picture}
\caption{Magnetization versus time curve for $L=10, T = 0.11 T_c$, 
  $h/J = 0.25$. Equilibrium probability distribution was estimated using the
Wang-Landau method up to a correction factor log($f$) = $1.25 \times 10^{-8}$.
In agreement with previous works~\cite{rikvold,tomita}, the switching duration
is approximately equal to the lifetime. Error bars are much smaller than 
thickness of the line.}
\label{fig:mag}
\end{figure}
\begin{figure}
\begin{picture}(0,210)
\put(50,0){\scalebox{0.33}{\includegraphics{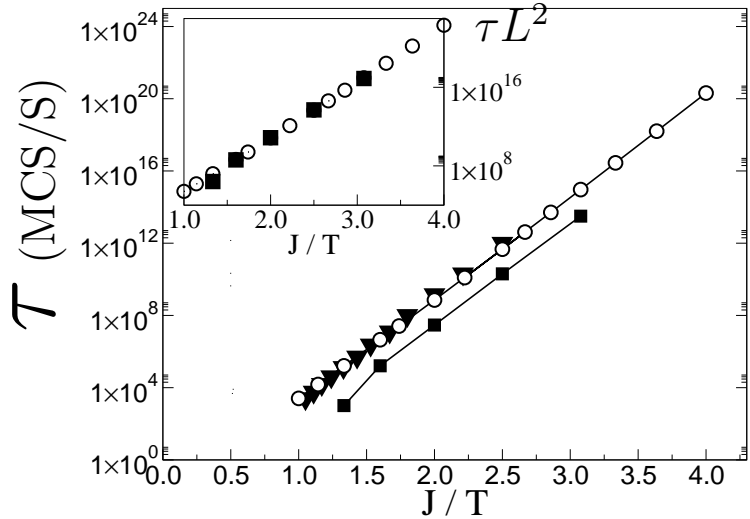}}}
\put(25,70){\scalebox{1.5}{\rotatebox{90}{\mbox{\huge $\tau$} (MCS/S)}}}
\put(200,180){\scalebox{1.5}{\mbox{$\tau L^2$ }}}
\end{picture}
\caption{Plot of switching time $\tau$ at $h/J = 0.75$, which shows very good 
agreement with previous studies by Novotny~\cite{novotny1}. Lattice sizes are 
$L = 10$ (circles) and $L = 50$ (squares). Triangles show the results reported
by Novotny~\cite{novotny1} for $L = 10$ lattice.}
\label{fig:tau}
\end{figure}
\begin{eqnarray}
\vec P(t)  & = & \alpha_{0} \vec{v}_0 + 
\sum_{i=1}^{N} \alpha_i \vec v_i \mbox{exp}(- |\lambda_i| t) \\
& = &  \vec{P}^{eq} + 
\sum_{i=1}^{N} \alpha_i \vec v_i \mbox{exp}(- |\lambda_i| t)
\label{eq:pt1}
\end{eqnarray}
where $\alpha_0$ and $\vec{v}_0$ are the coefficients and eigenvector
corresponding to the largest eigenvalue (which is zero), and
$\alpha_0 \vec{v}_0 = \vec{P}^{eq}$. Hence,
\begin{equation}
\lim_{t \rightarrow \infty} \vec{P}(t) = \vec{P}^{eq}
\end{equation}

Fig. \ref{fig:prob} shows the time evolution of the probability distribution
function for $L=10, T=0.44T_c$ and $h=0$.
The magnetization was set to $-N$ at $t=0$ and allowed to evolve
via the master equation. At $t=10$, probability distribution falls off 
exponentially from near 1 at $M=-100$ to exp(-225) at $M=100$. At $t=10^{10}$, the 
probability distribution at $M=100$ increases and we begin to see two peaks, but
the peak at $M=100$ is 20 orders of magnitude smaller. At $t=10^{25}$, the
probability distribution is almost symmetric and indistinguishable from the
equilibrium probability distribution. Spikes indicated with arrows
originate from special band structures of the Ising model
(Fig. \ref{fig:band}). These band structures are local energy minimums, any
single spin flip from these structures resulted in an increase of energy.
Fig. \ref{fig:band} also shows the domain of accessible states for $L = 10$,
and arrows indicate band structures states. They have equal
spacing in magnetization of $\Delta M=2L$ and their energies
are $E_g + 4L$, where $E_g$ is the ground state energy.
We have demonstrated our calculation of the probability distribution with
zero external field, because the spikes in the probability distribution 
appear most prominently with $h=0$. 

We are also interested in studying magnetic switching in an external
field where a direct comparison with previous work is available.
Fig. \ref{fig:mag} shows reversal of magnetization
in an external field at $T= 0.11 T_c$ and $h/J=0.25$ for $L=10$ lattice. 
In agreement with previous works~\cite{rikvold,tomita} the average switching
duration is approximately equal to the lifetime. 

\begin{table}
\begin{centering}
\begin{tabular}{|c|c|c|c|c|} \hline
$J/T$  & Rates & log($f$) values & $\tau$ (MCS/S) & $\Delta \tau$ \\ \hline
2.67 & Glauber    & $1.70 \times 10^{-4}$ & $4.1 \times 10^{12}$ & $5.8 \times 10^{11}$ \\ \hline
2.67 & Glauber    & $1.25 \times 10^{-6}$ & $4.1 \times 10^{12}$ & $3.9 \times 10^{10}$ \\ \hline
2.67 & Glauber    & $1.25 \times 10^{-8}$ & $4.1 \times 10^{12}$ & $2.8 \times 10^{10}$ \\ \hline
2.67 & Metropolis & $1.70 \times 10^{-4}$ & $3.9 \times 10^{12}$ & $5.6 \times 10^{11}$ \\ \hline
2.67 & Metropolis & $1.25 \times 10^{-6}$ & $4.0 \times 10^{12}$ & $3.7 \times 10^{10}$ \\ \hline
2.67 & Metropolis & $1.25 \times 10^{-8}$ & $4.0 \times 10^{12}$ & $2.7 \times 10^{10}$ \\ \hline
1.00 & Glauber    & $1.70 \times 10^{-4}$ & $2538 $ & $304 $ \\ \hline
1.00 & Glauber    & $1.25 \times 10^{-6}$ & $2550 $ & $21 $ \\ \hline
1.00 & Glauber    & $1.25 \times 10^{-8}$ & $2556 $ & $13 $ \\ \hline
1.00 & Metropolis & $1.70 \times 10^{-4}$ & $1524 $ & $182$ \\ \hline
1.00 & Metropolis & $1.25 \times 10^{-6}$ & $1527 $ & $12 $ \\ \hline
1.00 & Metropolis & $1.25 \times 10^{-8}$ & $1531 $ & $7.8 $ \\ \hline
\end{tabular}
\caption{Comparing results using Glauber and Metropolis transition rates
and using different estimates of equilibrium distributions. ``log($f$) values" 
given in the third column are the values of correction factors in Wang-Landau 
method~\cite{wang} on $L=10$ lattice with $h/J = 0.75$.}
\label{tbl:err}
\end{centering}
\end{table}%

The temperature dependence of the switching time can be studied by plotting
the switching times versus the inverse temperature at $h/J = 0.75$ 
(Fig. \ref{fig:tau}). 
Here we define the switching time as the time when the average magnetization
reaches zero ($\langle M(t) \rangle=0$).
The data for $L=10$ are compared with those by 
Novotny~\cite{novotny1}. We obtained very good agreement with previous works.
In the insert of Fig. \ref{fig:tau}, we plot the switching time $\times$ system
size as a function of the inverse temperature. The data for $L=10$ and $L=50$
collapse into a single curve, as expected in the single-droplet regime.
In contrast with the other method employed in Ref.~\cite{kolesik} 
where $\tau$ is calculated directly,
the present method calculates the entire probability distribution of the master
equation that allows us to calculate other quantities such as average magnetization.

There are two sources of approximation in our method; the first one is the 
approximation of transition rates using equilibrium distribution, the second 
one originated from uncertainty in the density of states estimated from 
Wang-Landau method. The effects of uncertainties on a $L=10$ lattice in a field
of $h/J=0.75$ are 
summarized in Table \ref{tbl:err}. 
In the Wang-Landau algorithm, the correction factor $f$ is used in the 
process of refining the density of states. The final value of $f$ is related
to the accuracy of the calculation, and smaller 
``log($f$)" values corresponds to more accurate equilibrium distributions. Our 
results suggest that running the Wang-Landau algorithm to log($f$) value of 
$1.25 \times 10^{-6}$ is sufficient to within our simulation precision. 
Table \ref{tbl:err}  also shows the dependence of switching time $\tau$ with 
Glauber and Metropolis rates. The Metropolis rates are given by,
\begin{eqnarray}
\omega( M|M') \approx \mbox{min} \left[1, \frac{P^{eq}(M)}{P^{eq}(M')} \right] & 
\mbox{ for } & |i-j|\leq 1
\end{eqnarray}
There is no significant difference in the 
switching time between Glauber and Metropolis transition rates for low 
temperature ($J/T = 2.67$). But they differ at high temperature because
the approximation to the transition rates becomes
less accurate at high temperatures.
The uncertainties in the switching time, $\Delta \tau$ was obtained from
several independent runs.

\section{Conclusion}

In conclusion, we derived an explicit expression for the magnetic
probability
distribution as a function of time. With the probability distribution, all 
dynamical information is available. And we used the information to obtain 
consistent results with previous works. Another advantage is 
that dynamical 
properties can be evaluated to arbitrary long time without requiring more 
computational resources. 
Lastly, we should mention that
our method is general and not restricted to discrete 
systems or any specific models. 
Future development of the present method should focus on calculation of
larger lattice sizes. Currently, lattice sizes are limited by the
efficiency of Wang-Landau algorithm to calculate joint density of states.

\section*{Acknowledgments}
We wish to thank J. S. Wang, M. A. Novotny, 
N. Kawashima, G. Brown and S. J. Mitchell for fruitful discussions.
This work is supported by the Japan Society for
Promotion of Science.

\section*{Appendix : Solutions to the Master Equation}

A general procedure to solve the master equation will be presented in this 
appendix. Firstly, we shall introduce the mathematical notation, 
the probability distribution function $P(M,t)$ will be denoted in a form of a
vector,
\begin{equation}
\left( \begin{array}{c}
       P(M=-N,\mbox{\hspace{6.5mm}}t) \\ P(M=-N+2,t) \\ 
       \vdots \\ P(M=+N\mbox{\hspace{6.5mm}},t) \end{array} \right) =
\left( \begin{array}{c}
       P_1(t) \\ P_2(t) \\ 
       \vdots \\ P_{N+1}(t) \end{array} \right)
\end{equation}
Note that for an Ising model with $N$ sites, with $N$ being an even number,
the number of possible 
magnetizations is $N+1$ with $M = \{-N, \cdots 0, \cdots N\}$. The master 
equation can be written in a matrix form,
\begin{equation}
\frac{d \vec P(t) }{dt} = {\mathcal A} \cdot \vec P(t)
\label{eq:masterMeq}
\end{equation}
where $\mathcal A$ is a $(N+1) \times (N+1)$ tridiagonal matrix with constant 
matrix elements.
Suppose $ {\mathcal S}^{-1} {\mathcal A}{\mathcal S} = {\mathcal D}$, where $\mathcal S$ is a 
similarity transform and $\mathcal D$ is a diagonal matrix. 
Then we have
\begin{eqnarray}
\frac{d \vec P(t) }{dt} & = & {\mathcal S} {\mathcal D} {\mathcal S}^{-1} \cdot \vec P(t) 
\label{eq:masterMeq1}
\end{eqnarray}
Defining a new vector $\vec{Q}(t) = {\mathcal S}^{-1} \vec{P}(t)$, we obtain
\begin{equation}
\frac{d \vec Q(t) }{dt} = {\mathcal D} \cdot \vec Q(t)
\label{eq:masterMeqQ}
\end{equation}
Since $\mathcal D$ is a diagonal matrix, Eq. (\ref{eq:masterMeqQ}) is easily solved
with $Q_i (t)  = C_i \mbox{ exp}(\lambda_i t)$, where $\lambda_i$ is the $i$th 
eigenvalue of $\mathcal A$. With $\vec P(t) = {\mathcal S} \vec Q(t)$, we have
\begin{equation}
\vec P(t) = \sum_i \alpha_i \vec{v}_i \mbox{ exp}(\lambda_i t)
\end{equation}
We write $\vec P$ in terms of the eigenvectors so that the constant 
coefficients $\alpha_i$ can be easily found by initial conditions,
\begin{equation}
\vec P(t=0) = \sum_i \alpha_i \vec{v}_i
\end{equation}


\begin{thebibliography}{99}
\bibitem{novotny}     M. A. Novotny, Phys. Rev. Lett. 74 (1995) 1.
\bibitem{bortz}       A. B. Bortz, M. H. Kalos, J. L. Lebowitz, J. Comput.  Phys. 17 (1975) 10.
\bibitem{kolesik}     M. Kolesik, M. A. Novotny, P. A. Rikvold, Phys. Rev. Lett. 80 (1998) 3384.
\bibitem{novotny1}    M. A. Novotny, Int. J. Mod. Phys. C 10 (1999) 1483.
\bibitem{munoz}       J. D. Munoz, M. A. Novotny, S. J. Mitchell, Phys. Rev. E 67 (2003) 26101.
\bibitem{brown}       G. Brown, M. A. Novotny, Per Arne Rikvold, J. Appl. Phys. 93 (2003) 6817.
\bibitem{hinzke}      D. Hinzke, U. Nowak, Phys. Rev. B 58 (1998) 265.
\bibitem{nowak1}      U. Nowak, Ann. Rev. Comp. Phys. IX (2001) 105.
\bibitem{nowak2}      U. Nowak, D. Hinzke, J. Appl. Phys. 85 (1999) 4337.
\bibitem{dittrich}    R. Dittrich, T. Schrefl, D. Suess, W. Scholz, H. Forster, J. Appl. Phys. 93 (2003) 7405.
\bibitem{wangjs}      R. H. Swendsen, J. S. Wang, Phys. Rev. Lett. 58 (1987) 86.
\bibitem{ferrenberg1} A. F. Ferrenberg, R. H. Swendsen, Phys. Rev. Lett. 61 (1988) 2635.
\bibitem{wolff}       U. Wolff, Phys. Rev. Lett. 62 (1989) 361.
\bibitem{ferrenberg2} A. F. Ferrenberg, R. H. Swendsen, Phys. Rev. Lett. 63 (1989) 1195.
\bibitem{lee1}        J. Lee, J. M. Kosterlitz, Phys. Rev. Lett. 65 (1990) 137.
\bibitem{oliveira}    P. M. C. de Oliveira, T. J. P. Penna, H. J. Herrmann, Eur. Phys. J. B 1 (1998) 205.
\bibitem{yamaguchi1}  C. Yamaguchi, N. Kawashima, Phys. Rev. E 65 (2002) 056710.
\bibitem{yamaguchi2}  C. Yamaguchi, N. Kawashima, Y. Okabe, Phys. Rev. E 66 (2002) 036704.
\bibitem{wang}        F. Wang, D. P. Landau, Phys. Rev. Lett. 86 (2001) 2050.
\bibitem{wang1}       F. Wang, D. P. Landau, Phys. Rev. E 64 (2001) 056101.
\bibitem{lee}         J. Lee, M. A. Novotny, P. A. Rikvold, Phys. Rev. E 52 (1995) 356.
\bibitem{swendsen}    J. S. Wang, R. H. Swendsen, J. Stat. Phys. 106 (2002) 245.
\bibitem{langer1}     J. S. Langer, Phys. Rev. Lett. 21 (1968) 973.
\bibitem{langer2}     J. S. Langer, Ann. Phys. 54 (1969) 258.
\bibitem{rikvold}     P. A. Rikvold, H. Tomita, S. Miyashita, S. W. Sides, Phys. Rev. E 49 (1994) 5080.
\bibitem{tomita}      H. Tomita, S. Miyashita, Phys. Rev. B 46 (1992) 8886.
\bibitem{ince}        Chapter VI, E.L. Ince, {\it Ordinary Differential Equations}, Dover Publications.  
\end{thebibliography}
\end{document}